\documentclass{ws-mpla}

\begin{document}

\markboth{S. Odaka}
{Simulation of $Z$ boson $p_{T}$ spectrum at Tevatron by leading-order event generators}

\catchline{}{}{}{}{}

\title{SIMULATION OF $Z$ BOSON $p_{T}$ SPECTRUM AT TEVATRON 
BY LEADING-ORDER EVENT GENERATORS}

\author{\footnotesize SHIGERU ODAKA}

\address{High Energy Accelerator Research Organization (KEK)\\
1-1 Oho, Tsukuba, Ibaraki 305-0801, Japan\\
shigeru.odaka@kek.jp}

\maketitle

\pub{Received (Day Month Year)}{Revised (Day Month Year)}

\begin{abstract}
We show that the transverse momentum ($p_{T}$) spectrum of $Z$ boson production 
measured at Fermilab Tevatron can be well reproduced by leading-order event generators 
if $Z$ + 1 jet processes are included with a proper solution for the double-count problem 
and if the parton shower (PS) branch kinematics are defined appropriately.
The choice of the PS evolution variable does not definitely determine the low-$p_{T}$ behavior.
Our new event generator employing the limited leading-log (LLL) subtraction 
and a built-in leading-log PS reproduces the spectrum very well, 
not only in large $p_{T}$ regions but also at low $p_{T}$ down to $p_{T}$ = 0.

\end{abstract}

\ccode{PACS Nos.: 12.38.-t, 13.85.Qk}

\section{Introduction}

$Z$ boson production in hadron collisions is one of the ideal places 
for testing quantum chromodynamics (QCD), 
because we can focus on its role in the initial state 
if the production is tagged by the leptonic decays of $Z$ bosons. 
Leptonic tagging allows unambiguous measurements of the production kinematics
in a wide range of the transverse momentum ($p_{T}$) of $Z$ bosons 
with respect to the colliding beam direction.
At high $p_{T}$, typically around the $Z$ boson mass ($m_{Z}$) or higher, 
the interaction is expected to be well described by a first order perturbative calculation 
of QCD including the production of an additional energetic quark or gluon.
On the other hand, at low $p_{T}$ ($\ll m_{Z}$), multiple radiation effects 
and non-perturbative effects of QCD are expected to determine the production kinematics.

Monte Carlo (MC) event generators used for evaluating the detection efficiency 
and acceptance of experiments
are desired to simulate the phenomena continuously in the entire kinematical range 
including the above qualitatively different $p_{T}$ regions.
The performance of MC event generators may depend on the actual implementation  
because the interpolation between the two $p_{T}$ regions is not trivial 
and the introduction of models is necessary to simulate low-$p_{T}$ phenomena.
In this paper, we discuss the performance of existing MC event generators 
and a new MC event generator that we have developed, 
by comparing their predictions with experimental measurements 
on the $p_{T}$ spectrum of $Z$ bosons 
at Fermilab Tevatron\cite{Affolder:1999jh,Abbott:1999wk,Abazov:2007nt}: 
$p\bar{p}$ collisions at center-of-mass (cm) energies ($\sqrt{s}$) of 1.8 and 1.96 TeV.
Since the main use of MC event generators is for the acceptance and efficiency evaluations, 
experimentalists are mainly interested in the accuracy of relative spectra 
rather than the absolute value of the interaction cross sections.
We therefore focus on the relative shape of the $p_{T}$ distribution in this paper.

The cross sections of hard interactions in hadron collisions, such as $Z$ boson production, 
are usually evaluated by separating hard interaction parts initiated by constituent partons 
(light quarks and gluons) 
from soft/collinear divergent contributions of additional parton radiations, 
in order to improve the convergence of perturbation.
The soft/collinear contributions can be factorized and added to all orders 
of the coupling constant ($\alpha_{s}$) to give finite results.
The results are provided in the form of parton distribution functions (PDF) 
which depend on the energy scale (factorization scale) of the hard interactions.

MC event generators also employ the above approach.
The hard interaction events are generated according to fixed-order matrix elements (ME), 
while the soft/collinear contributions are simulated with parton showers (PS).
MEs are usually at the leading order (LO), 
and the implementation of PS is limited to the leading-logarithmic (LL) approximation 
for technical reasons, 
although PDFs including sub-leading contributions are already widely used.
There are two kinds of parton showers: the spcelike PS to be applied to the initial-state partons 
and the timelike PS in the final state.
Hereafter we focus on the initial-state PS 
since we discuss about the $Z$ boson production tagged by leptonic decays.

The initial-state PS should in principle give the same answer as PDF 
for the momentum distribution of partons along the beam direction 
if the approximation level is identical.
In addition to the longitudinal properties, 
PS simulates transverse momenta of the radiations 
in order to provide an exclusive simulation of interactions.
We need to introduce a certain model for this simulation, 
because PDFs are evaluated at an infinite momentum limit 
while PS simulations have to be constructed in a finite momentum frame.
The main issue in the model is to define the relation between variables 
in the infinite-momentum frame and kinematical variables in the finite-momentum frame.
Thus, the $p_{T}$ spectra may depend on the introduced model.

The transverse activities of QCD radiations result in a finite transverse recoil 
of hard interaction events, 
and produce additional hadronic activities visible in detectors.
Since these effects alter the acceptance and detection efficiency of measurements, 
MC event generators implementing PS are indispensable tools 
for hadron collision experiments.
Because any inaccuracy in the simulation may deteriorate the measurement precision, 
MC event generators are required to reproduce actual phenomena as precisely as possible 
in the whole kinematical region.

The transverse recoil can also be evaluated analytically 
by the resummation\cite{Balazs:1997xd,Bozzi:2008bb}.
The resummation studies usually include next-to-leading logarithmic (NLL) 
or further sub-leading  contributions, 
and provide predictions in good agreement with the measured $Z$ boson 
$p_{T}$ spectrum\cite{Affolder:1999jh,Abbott:1999wk,Abazov:2007nt,Balazs:2000sz}.
On the other hand, 
it is known for long time that the most widely-used event generator, 
PYTHIA\cite{Sjostrand:1993yb}, 
shows substantially softer $p_{T}$ spectra of $Z$ bosons.
The deviation is significant at low $p_{T}$ where PS is expected to play a dominant role.
The authors in a previous study\cite{Balazs:2000sz} suggested 
that the deviation might be partly due to a sub-leading effect missing in the PS simulation.
The problem must be serious if the deviation is actually due to the lack 
of sub-leading contributions.
In order to make the situation clearer, 
we first re-examine the performance of existing MC event generators.

\section{PYTHIA and HERWIG}

\begin{figure}[t]
\centerline{\psfig{file=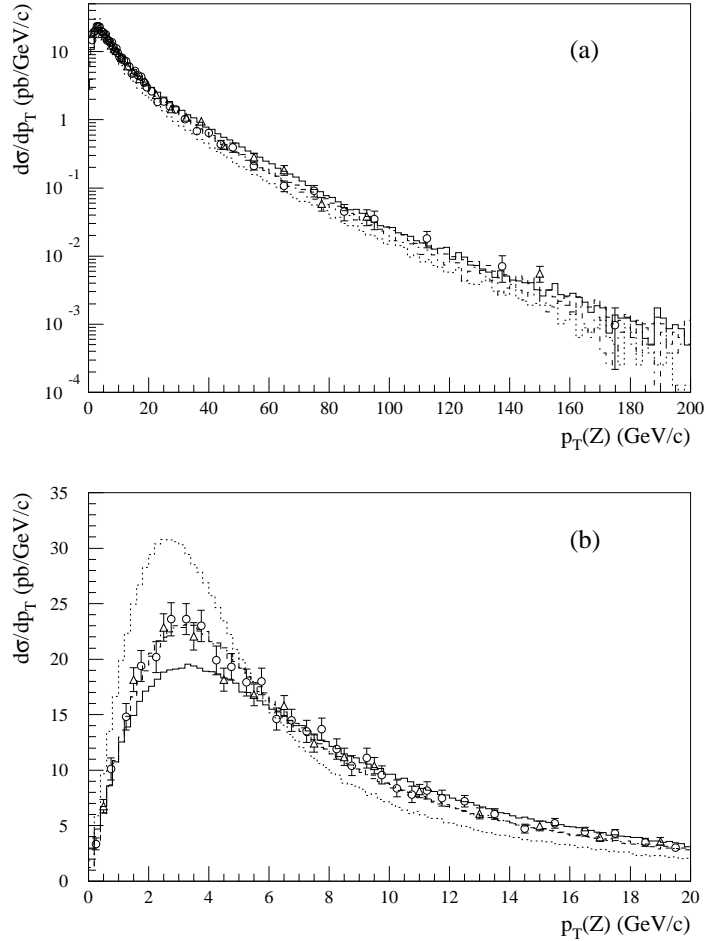,width=100mm}}
\vspace*{8pt}
\caption{The $p_{T}$ spectrum of $Z$ bosons at Tevatron Run 1.
PYTHIA and HERWIG simulations are compared with the CDF (circles) 
and D0 (triangles) data. 
The histograms show the HERWIG result (solid), and the PYTHIA results 
with the new PS (dashed) and the old PS (dotted).
The CDF Tune-AW result is also shown with a dot-dashed histogram, 
but hard to be distinguished from the PYTHIA new PS result.
The D0 data and the simulations are normalized to the total cross section of CDF (248 pb).
\protect\label{fig:pyhw}}
\end{figure}

The simulations by the most popular LO event generators, PYTHIA and HERWIG, 
are compared in Fig.~\ref{fig:pyhw} with the experimental data 
for the $Z \rightarrow e^{+}e^{-}$ channel
measured by CDF\cite{Affolder:1999jh} and D0\cite{Abbott:1999wk} 
at Tevatron Run 1 ($\sqrt{s}$ = 1.8 TeV).
The D0 data (triangles) are multiplied by a factor of 1.12 
to compensate for the difference in the total cross section
with respect to the CDF data (circles)\cite{Affolder:1999jh,Abbott:1999tt}.
Histograms show the results of PYTHIA 6.419\cite{Sjostrand:2006za} 
and HERWIG 6.510\cite{Corcella:2000bw,Corcella:2002jc}.
PYTHIA results are shown for two different PS models: the old PS (default, MSTP(81) = 1) 
and the new PS (MSTP(81) = 21).
All the other parameters are unchanged from the default.
The HERWIG result is also obtained using its default setting, 
but a so-called primordial $k_{T}$ effect is added by smearing the generated events 
with an additional $p_{T}$ that is randomly generated according to 
the two-dimensional Gaussian distribution having one standard deviation of 2.0 GeV/$c$.
This value is smaller than the PYTHIA default by a factor of $\sqrt{2}$, 
because the same amount of smearing is applied to each beam in PYTHIA.
We expect that the primordial $k_{T}$ should include perturbative effects in a soft region 
which PS does not cover, as well as non-perturbative effects inside hadrons.
Therefore, it may depend on the PS implemented in the simulation.

According to the signal definition of CDF, 
an invariant mass cut ($66 < m_{Z} < 116$ GeV/$c^{2}$) is applied to the generated events 
in all the simulations, 
although the specific cut values are not important in the present study. 
The choice of PDF is not important, too.
Though the used PDFs are rather old, 
we observe no significant change even when we replace them with a recent one (CTEQ6L1).
All the results are normalized to the total cross section measured by CDF 
in order to compare the relative spectra.

The so-called ME correction is applied in all the simulations, 
where the maximum energy scale of PS is set to a very large value, 
and the generated events are reweighted so that the frequency of the hardest radiation
matches the prediction from the matrix element (ME) calculation including one additional jet 
(a parton in the final state). 
Thus, the $p_{T}$ spectrum should be equal to the prediction from the $Z$ + 1 jet ME 
at high $p_{T}$ if the behavior of PS is correctly known.
The jet radiation is naturally suppressed at low $p_{T}$ as a result of higher-order effects 
since it is generated by PS.
This is one of the solutions for the double-count problem: 
a problem that arises when we naively apply the $Z$ + 1 jet ME for event generation, 
where the radiation effects are doubly counted by the ME and PS.

First, look at the distribution at high $p_{T}$ ($\gtrsim 30$ GeV/$c$) in Fig.~\ref{fig:pyhw}a.
Although the PYTHIA old PS simulation (dotted) gives a continuously smaller prediction, 
all the simulations reasonably reproduce the experimental data, 
at least in their $p_{T}$ dependence.
This shows that the ME correction works well, and that the lowest-order ME calculation 
for one additional jet is sufficient for reproducing the $p_{T}$ dependence at high $p_{T}$.
The predictions of PYTHIA new PS (dashed) and HERWIG (solid) are close to each other, 
except in the medium $p_{T}$ region ($40-80$ GeV/$c$), 
where the HERWIG result shows a small enhancement.

The difference between the three simulations is obvious 
in the low-$p_{T}$ region ($<$ 20 GeV/$c$) shown in Fig.~\ref{fig:pyhw}b.
PYTHIA new PS reproduces the data very well, 
while PYTHIA old PS gives a higher peak and HERWIG gives a lower peak.
The peak positions ($\sim$3 GeV/$c$) are almost identical because they are, 
at least in the PYTHIA simulations, predominantly determined by the primordial $k_{T}$. 
On the other hand, the peak height is determined by the distribution in off-peak regions 
($\gtrsim 6$ GeV/$c$) since the simulations are normalized to the total yield. 
The HERWIG result shows some enhancement in off-peak low-$p_{T}$ 
and medium-$p_{T}$ regions.
The peak height reduces to compensate for the enhancement.
PYTHIA old PS gives continuously lower predictions in the off-peak regions.
As a result, the peak becomes very high.
The PYTHIA simulation in the previous study\cite{Balazs:2000sz} 
corresponds to this PYTHIA old PS simulation.
The above result is consistent with the previous observation.

The principal difference between the three simulations is 
in the choice of the evolution (ordering) variable in PS.
The virtuality ($Q^{2}$) is chosen in PYTHIA old PS,
while $p_{T}$ is used in the new PS, and an angle-based variable in HERWIG.
Do the above results imply that the $p_{T}$ spectrum depends on the choice of 
the evolution variable, and that $p_{T}$ is the best choice among them?

We have illustrated another simulation result based on PYTHIA old PS 
in Fig.~\ref{fig:pyhw} (dot-dashed histogram).
It comes from the so-called CDF Tune-AW\cite{Field:2008zz}, 
where parameters in PYTHIA are tuned by CDF 
so that the simulation well reproduces their data.
We have used an older version of PYTHIA (6.212) for this simulation 
because the tuning is sensitive to it.
It is difficult to distinguish this result from the PYTHIA new PS simulation; 
namely, the result reproduces the data very well.

Though the result is good, 
this tuning includes an unnatural setting of a parameter: PARP(64) = 0.2.
The squared energy scale used for evaluating $\alpha_{s}$ in PS 
is multiplied by this parameter. 
If we reset it to the default value ($=1.0$), 
we obtain a result similar to the PYTHIA old PS simulation shown in Fig.~\ref{fig:pyhw}.
Since the coupling strength is significantly changed, 
this tuning explicitly violates the consistency between PS and PDF.
The success of this tuning may imply that the success of PYTHIA new PS may also be 
attributed to some tuning that is specific to the referred measurements.
An independent test is necessary in order to clarify such a suspicion 
and to answer the question concerning the choice of the evolution variable.

\section{GR@PPA simulation}

As reported in our previous paper\cite{Odaka:2007gu}, 
we have developed a technique to consistently merge ME calculations for 
$W$ + 0 jet and $W$ + 1 jet production processes in hadron collisions 
by employing the limited leading-log (LLL) subtraction\cite{Kurihara:2002ne} 
together with a built-in leading-log (LL) PS.
This is another solution (a matching method) for the double-count problem.
We have recently imported this technique 
into the GR@PPA event generator\cite{Tsuno:2006cu}, 
where MEs have been generated using the GRACE system\cite{Ishikawa:1993qr}, 
and the cross-section integration and event unweighting are automatically carried out 
within the framework of BASES/SPRING\cite{Kawabata:1985yt,Kawabata:1995th}.
As a result, it is now rather easy to apply our matching method to other processes, 
because many processes are already implemented in GR@PPA.

We have first applied the matching method to $Z$ boson production 
and simulated the process in the Tevatron Run 1 condition.
The simulation has been carried out by setting the factorization scale ($\mu_{F}$) 
and the renormalization scale ($\mu_{R}$) 
to be equal to the $Z$ boson mass ($m_{Z}$ = 91.19 GeV) as the default.
In our matching method, 
the divergent leading-logarithmic (LL) components are numerically subtracted 
from the $Z$ + 1 jet MEs when $Q^{2}$ ($= -t$) of the jet is smaller than $\mu_{F}^{2}$; 
consequently, the remaining $Z$ + 1 jet cross section becomes finite. 
Though a $p_{T}$ cut ($p_{T} >$ 1 GeV/$c$) is applied to the jet for numerical stability, 
its effect is negligible since the differential cross section converges to zero 
as $p_{T} \rightarrow 0$ after the subtraction.

Since the subtraction is unphysical, 
we may have negative cross sections in some phase space, 
leading to the generation of events having a negative weight.
Note that, since the unweighting is automatically done by SPRING, 
the generated events always have a weight of $+1$ or $-1$.
Physical distributions can be obtained by subtracting the number of negative-weight events 
from the number of events having a weight of $+1$ in each histogram bin.
Though this subtraction deteriorates the statistical accuracy, 
it is not a serious problem because the overall fraction of the negative-weight 
events is only 2\% in the present simulation.

The subtracted LLL components are restored by applying a PS to the $Z$ + 0 jet simulation.
Thus, the maximum energy scale of PS ($\mu_{PS}$) must be equal to $\mu_{F}$ 
in our method.
Our PS employs a forward evolution technique\cite{Kurihara:2002ne} 
based on the Sudakov form factor at the leading order, which is expressed as
\begin{equation}\label{sudakov}
	S(Q_{1}^{2}, Q_{2}^{2}) = \exp\left[ - \int_{Q_{1}^{2}}^{Q_{2}^{2}}
	{dQ^{2} \over Q^{2}} \int_{\epsilon}^{1-\epsilon} dz \  
	{\alpha_{s}(Q^{2}) \over 2\pi}\ P(z) \right] ,
\end{equation}
where $P(z)$ symbolically represents the leading-order splitting functions 
summed over all possible branches. 
The PS branches are, therefore, ordered in $Q^{2}$.
The parameter $\epsilon$ cuts off the divergences in the splitting functions.
We take $\epsilon = 10^{-6}$ as the default.
Physical properties such as cross sections and the $p_{T}$ distribution of $Z$ 
do not depend on this cutoff if it is sufficiently small.
We have confirmed the independence by changing the cutoff value down to $10^{-10}$.

In our simulation, PS replaces the QCD evolution in PDF.
A PDF is used for determining the initial condition at a small $Q^{2}$ ($Q_{0}^{2}$).
We set $Q_{0}$ = 4.6 GeV and use CTEQ6L1\cite{Pumplin:2002vw} 
for the initial PDF. 
Light quarks up to $b$ and gluons are taken as the constituents.
This simulation gives results close to the ones obtained by the conventional method 
that uses PDF for the evolution, 
with a precision at the level of 1\% in the total cross section.

The theoretical basis of our PS is in principle same as that of PYTHIA old PS; 
namely, the approximation is at the LL order and the branches are ordered in $Q^{2}$.
The difference is in the model of branch kinematics.
PYTHIA old PS identifies $Q^{2}$ as the virtuality of evolving partons 
and the $z$ parameter as $z = \hat{s}'/\hat{s}$, 
where $\hat{s}$ and $\hat{s}'$ are the squared cm energies of the colliding partons 
before and after branching, respectively. 
The momenta of partons are calculated from the energy-momentum conservation 
based on these definitions\cite{Sjostrand:2006za}.
The $p_{T}$ of the branch is, therefore, determined indirectly in this calculation.
On the other hand, in our PS, while the definition of $z$ is the same,
the $p_{T}$ of each branch with respect to the colliding parton direction is 
"prefixed"\cite{Odaka:2007gu} according to the relation, 
\begin{equation}\label{ptdef}
  p_{T}^{2} = (1-z)Q^2 .
\end{equation}
This is the simplest relation derived from a massless approximation 
where the incoming partons and the radiation are assumed to be massless.
Four-momenta of the radiation and the recoil parton are determined 
from the $p_{T}$ and $z$ values and the four-momenta of the partons before branching 
(the $p_{T}$-prefixed branch kinematics).
As a result, $Q^{2}$ is no longer identical to the virtuality of the evolving partons.
Though not often, $p_{T}$ given by Eq.~(\ref{ptdef}) may exceed 
the kinematically allowed maximum.
The $p_{T}$ value is set to the allowed maximum in such cases.
In addition, the kinematics determination fails in about 0.1\% of the events 
because the calculated virtuality of the recoil parton becomes too large.
The event generation is retried in this case.

The PS is also applied to the LLL-subtracted $Z$ + 1 jet events with the same 
energy scale setting.
Therefore, it never generates secondary radiations harder than $\mu_{F}$.
In our previous study\cite{Odaka:2007gu}, 
we required that the $p_{T}$ of PS branches should never exceed that of the jet in ME.
This requirement is not applied in the present study 
because there is no reason to require such an ordering to non-logarithmic components 
remaining at $Q^{2} < \mu_{F}^{2}$.

The generated events are stored as a simple dump of the LHA event record\cite{Boos:2001cv} 
and fed to PYTHIA 6.419. 
The energy scale in the event record is set to 4.6 GeV, {\it i.e.}, the $Q_{0}$ value of our PS.
PYTHIA adds PS at lower energy scales and applies further hadronization and decay simulations.
Among them the most important is the primordial $k_{T}$ in the present study.
Applied is the PYTHIA 6.419 default, {\it i.e.}, $\sigma(k_{T})$ = 2.0 GeV/$c$.
The default setting in PYTHIA remains unchanged, except for the parameter PARP(67).
This is a scaling factor for $\mu_{PS}^2$ of the initial-state PS. 
The default value is 4.0; namely, the energy scale given in the event record is doubled.
Since we do not want the energy scale to increase, we reset it to 1.0.
However, these details, except for the primordial $k_{T}$, are not important in the present study.

\begin{figure}[t]
\centerline{\psfig{file=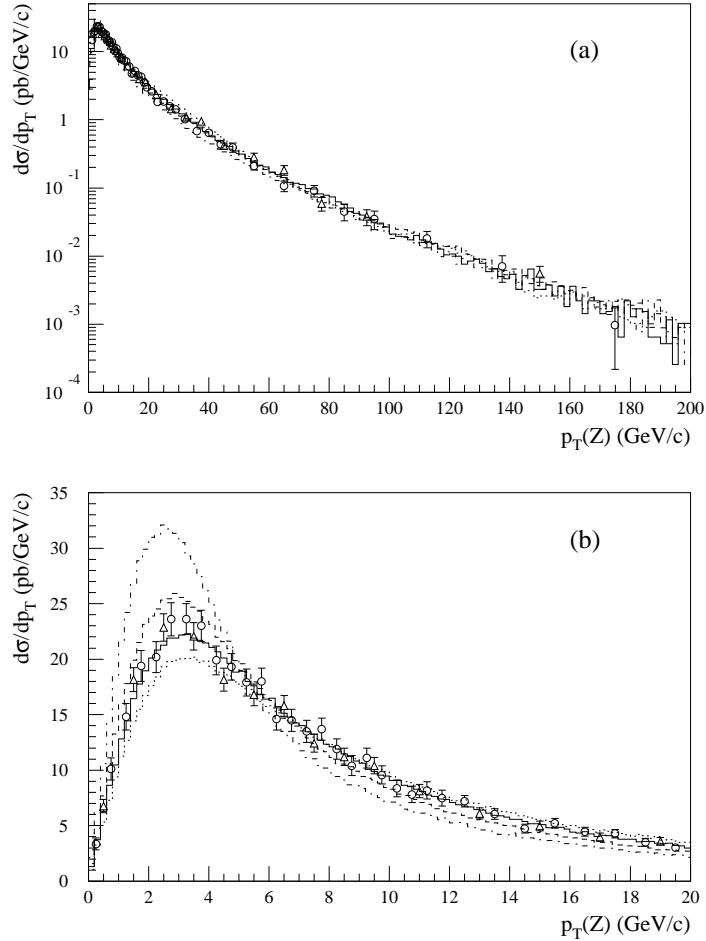,width=100mm}}
\vspace*{8pt}
\caption{Same as Fig.~\ref{fig:pyhw}, 
but compared are the simulations by GR@PPA. 
The histograms show the simulation results 
with $\mu_{F}/m_{Z}=1.0$ (solid), 0.5 (dashed), and 1.5 (dotted).
The dot-dashed histogram shows the result with an optional $p_{T}$ definition 
of the PS branch, $p_{T}^{2}=(1-z)^{2}Q^{2}$.
\protect\label{fig:grappa}}
\end{figure}

The simulation result is shown in Fig.~\ref{fig:grappa} (solid histogram).
The total yield is again normalized to the CDF cross section.
We can observe that the result reproduces the experimental data very well.
It is not surprising that good agreement is observed in the high-$p_{T}$ regions 
shown in Fig.~\ref{fig:grappa}a, 
since it is the purpose of our matching method.
On the other hand, the agreement at low $p_{T}$ is surprising 
because we have not considered the low-$p_{T}$ behavior.
Note that the contribution of the LLL-subtracted $Z$ + 1 jet is negligible 
in the low-$p_{T}$ region shown in Fig.~\ref{fig:grappa}b.

Our matching method has an explicit dependence on $\mu_{F}$ 
in both non-radiative (0 jet) and radiative (1 jet) processes.
They reasonably cancel each other.
As a result, the simulations show good stability against the variation 
in $\mu_{F}$\cite{Odaka:2007gu}.
The remaining dependence shows the prediction uncertainty of our simulations.
We show the results with extreme choices of $\mu_{F}/m_{Z} = 0.5$ (dashed) 
and $1.5$ (dotted) in Fig.~\ref{fig:grappa}, where, though not important here, 
the renormalization scale ($\mu_{R}$) is set to be equal to $\mu_{F}$ 
in order to achieve the matching in the radiation probability at the boundary.
We can see that the variation in the simulation results is small, 
and the experimental results are well within this range.
Therefore, the $\mu_{F}$ value can be "tuned" if a more precise simulation is required, 
for instance, for acceptance and/or efficiency studies. 

We have introduced $p_{T}$ definitions other than Eq.~(\ref{ptdef}) 
in previous reports\cite{Odaka:2007gu,Bern:2008ef}: 
\begin{equation}\label{ptdef1}
  p_{T}^{2} = (1-z)^{2}Q^{2}
\end{equation}
and
\begin{equation}\label{ptdef2}
  p_{T}^{2} = (1-z-Q^{2}/{\hat s})Q^{2} .
\end{equation}
We have obtained Eq.~(\ref{ptdef1}) for the PYTHIA old PS kinematics at a certain limit, 
and shown that our PS with this relation well reproduces the $p_{T}$ spectrum 
of the old PS\cite{Odaka:2007gu}.
The dot-dashed histogram in Fig.~\ref{fig:grappa} illustrates the result 
for the $Z$ boson production.
In fact, it is quite similar to the old PS prediction in Fig.~\ref{fig:pyhw}, 
and fails in reproducing the experimental data.
Equation~(\ref{ptdef2}) can be derived from massless kinematics 
without taking the $Q^{2}/{\hat s} \rightarrow 0$ limit.
Though Eq.~(\ref{ptdef2}) may be better at high $p_{T}$\cite{Bern:2008ef}, 
it cannot give a sufficient distribution in off-peak regions, 
resulting in a soft low-$p_{T}$ spectrum such that obtained with Eq.~(\ref{ptdef1}) 
and failing in reproducing the data.
The choice of the PS kinematics model was postponed 
in our previous paper\cite{Odaka:2007gu}.
We can now conclude that the $p_{T}$-prefixed kinematics with Eq.~(\ref{ptdef}) 
is the best choice among the options that we have considered.

\section{Discussions}

\begin{figure}[t]
\centerline{\psfig{file=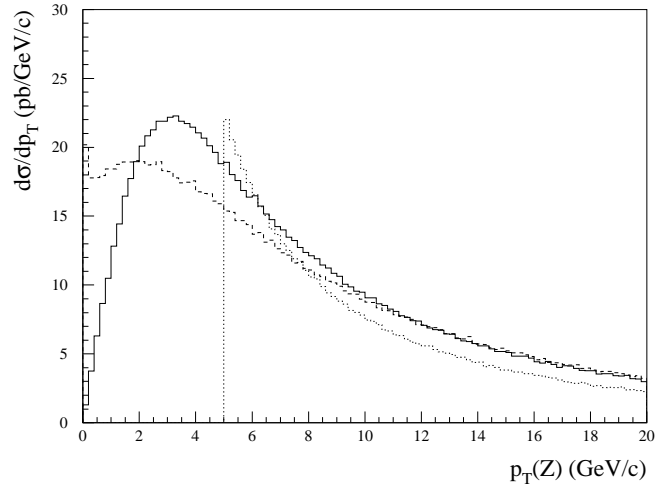,width=100mm}}
\vspace*{8pt}
\caption{The $p_{T}$ spectrum of $Z$ bosons 
simulated by GR@PPA in the Tevatron Run-1 condition.
The spectrum of $Z$ bosons generated by GR@PPA (dashed) is compared with 
the final result (solid) after applying the PYTHIA simulation.
The dotted histogram shows the spectrum derived from the $Z$ + 1 jet matrix element 
with a minimum  $p_{T}$ requirement of 5 GeV/$c$.
\protect\label{fig:low-pt}}
\end{figure}

In the GR@PPA event generation, the hard interaction and a hard part ($Q > 4.6$ GeV) of PS 
are simulated by GR@PPA. 
The PYTHIA simulation applied to the generated events adds PS at smaller $Q$ values  
and further interactions at lower energy scales.
The $p_{T}$ distributions of $Z$ bosons before and after adding the PYTHIA simulation 
are compared in Fig.~\ref{fig:low-pt}.
Concerning the $Z$-boson $p_{T}$ spectrum, 
the most effective in the PYTHIA simulation is the application of the primordial $k_{T}$.
The addition of a finite $p_{T}$ forms a clear peak in the spectrum as shown in the figure. 
The peak position is determined by the average $k_{T}$ value 
that we set in the simulation.

On the other hand, the peak height is determined by the distribution in off-peak regions.
We can see in Fig.~\ref{fig:low-pt} that the spectrum above $p_{T} = 10$ GeV/$c$ is 
not significantly altered by the PYTHIA simulation.
The spectrum in the off-peak region is predominantly determined by the simulation in GR@PPA.
We have also shown in Fig.~\ref{fig:low-pt} the $p_{T}$ spectrum derived from 
the matrix element (ME) for the $Z$ + 1 jet production.
The simulation result is apparently different from this ME prediction. 
Namely, the multiple radiation effect simulated by PS is significant 
in this $p_{T}$ region.
These are the reasons why the PS branching model applied in GR@PPA plays 
an important role to determine the low-$p_{T}$ spectrum.

The GR@PPA simulation results show that
LO MC event generators employing LL PS can reproduce the measured $Z$ boson 
$p_{T}$ spectrum very well 
if we adopt an appropriate PS branch model. 
The softer $p_{T}$ spectrum in the PYTHIA old PS simulation must be due to 
an inappropriate choice of the model, 
and sub-leading contributions missing in the PS simulation seem to be still insignificant. 
It should be emphasized that the present GR@PPA simulation is strictly subject to 
primitive theoretical arguments. 
It does not contain any adjustable parameter that may alter physical spectra, 
except for the choice of Eq.~(\ref{ptdef}).

The above results also show that
the choice of the evolution variable does not definitely determine the low-$p_{T}$ behavior, 
but the effective definition of $p_{T}$ in the PS branch kinematics does.
In PYTHIA new PS, though $p_{T}$ is adopted as the evolution variable, 
the $p_{T}$ of PS branches is not identical to the $p_{T}$ 
in the evolution\cite{Sjostrand:2006za}.
The $p_{T}$ given in the evolution is converted to $Q^{2}$ according to Eq.~(\ref{ptdef}) 
to determine the branch kinematics in the same manner as the old PS, 
where $Q^{2}$ is identified as the virtuality of evolving partons.
However, since the branches are ordered in $p_{T}$,
there must be large $p_{T}$ and $Q^{2}$ gaps between hard branches. 
Therefore, the branch $p_{T}$ should not be very different from the evolution $p_{T}$
that preserves the relation in Eq.~(\ref{ptdef}).
It should be noted that the $p_{T}$ gaps do not depend on any arbitrary parameter,
because the cutoff in the Sudakov form factor is determined by kinematical constraints 
in this PS scheme. 
This must be the reason why PYTHIA new PS gives a good result.

Parton branches necessarily produce non-zero virtuality in recoil partons. 
This virtuality affects the kinematics of subsequent branches, 
leading to a deviation from Eq.~(\ref{ptdef}).
The factorization theory is constructed at the limit where this difficulty can be ignored. 
However, we cannot ignore it since PS has to conserve the energy and momentum exactly. 
What we can do is to minimize the effect of this virtuality.
One of the ways to minimize the effect is to choose a physical quantity such as $p_{T}$ 
as the evolution variable, as discussed in the above. 
We have adopted another solution in our PS, 
where we directly determine the $p_{T}$ according to Eq.~(\ref{ptdef}) 
by discarding the identity between the $Q^{2}$ and the virtuality. 
The virtuality effect remains only as a boundary condition, 
and results in a small but finite failure rate in the kinematics determination. 

\section{Tevatron Run 2 result}

\begin{figure}[t]
\centerline{\psfig{file=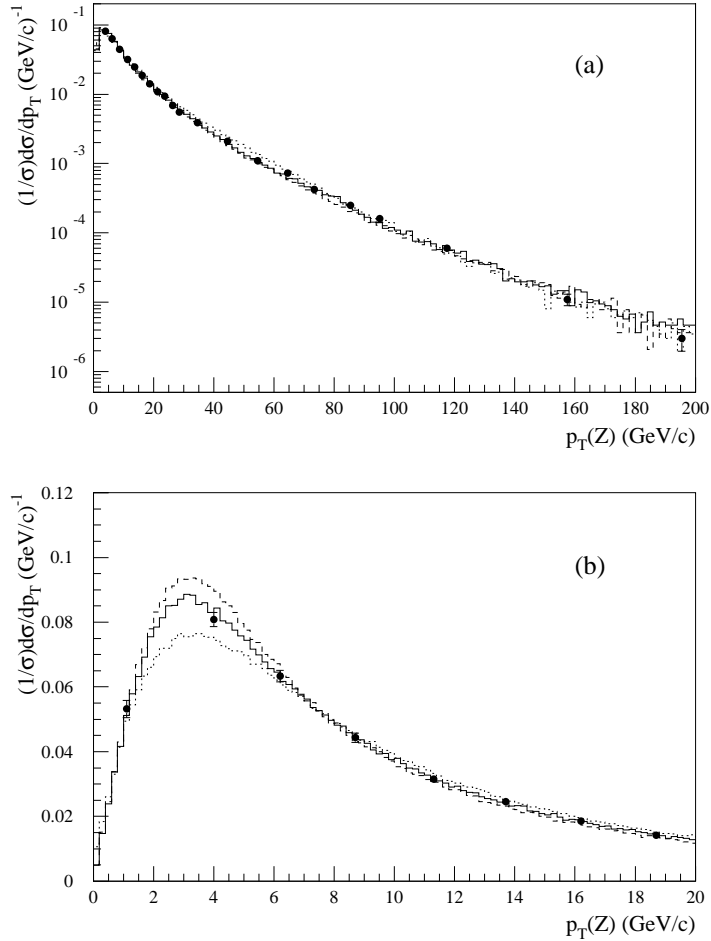,width=100mm}}
\vspace*{8pt}
\caption{The $p_{T}$ spectrum of $Z$ bosons at Tevatron Run 2. 
The simulations by GR@PPA (solid), PYTHIA new PS (dashed) and HERWIG (dotted) 
are compared with the D0 data.
\protect\label{fig:run2}}
\end{figure}

The D0 experiment has also measured the $Z$ boson $p_{T}$ spectrum 
at Tevatron Run 2 ($\sqrt{s} = 1.96$ GeV)\cite{Abazov:2007nt}.
The simulations by GR@PPA (solid), PYTHIA new PS (dashed) and HERWIG (dotted) 
performed in the Run 2 condition 
are compared with the published data in Fig.~\ref{fig:run2}.
The overall properties are the same as those observed for the Run 1 measurements. 
All the three simulations are nearly identical and in good agreement with the data 
in the high-$p_{T}$ region shown in Fig.~\ref{fig:run2}a, 
except for a small enhancement of HERWIG at medium $p_{T}$.
However, the D0 data show a slightly harder spectrum compared to the Run 1 data, 
especially compared to the CDF data. 
As a result, the HERWIG simulation is better and the PYTHIA new PS simulation 
is worse at low $p_{T}$, as we can see in Fig.~\ref{fig:run2}b. 
The GR@PPA simulation lies between the two.
Therefore, it is the best and reproduces the data very well. 
Unfortunately, the data points are too sparse in the low-$p_{T}$ peak region.
More data are desired for further discussions.

\section{Conclusion}

In conclusion, 
the $p_{T}$ spectrum of the $Z$ boson production at Tevatron can be reproduced 
with a good precision using leading order (LO) event generators 
employing leading-log (LL) PS.
The inclusion of radiative (1 jet) MEs with an appropriate solution for the double-count problem 
is necessary to give a good simulation at high $p_{T}$, 
while the definition of the PS branch kinematics is crucial for reproducing 
the low-$p_{T}$ spectrum.
Definitions which effectively give $p_{T}$ values close to the massless approximation 
provide good simulations at low $p_{T}$, 
irrespective of the choice of the evolution (ordering) variable in PS. 
Sub-leading contributions missing in PS seem to be still minor 
compared to the measurement accuracy.

The GR@PPA event generator that employs the LLL subtraction and a $Q^{2}$-ordered LL PS 
reproduces the experimental data very well in the entire $p_{T}$ range, 
if the $p_{T}$-prefixed kinematics with the simplest $p_{T}$ definition, Eq.~(\ref{ptdef}), 
is adopted for PS branches.
The adoption of an inappropriate branching model must be the reason 
why the PYTHIA old PS simulation does not well reproduce the experimental data. 

\section*{Acknowledgments}

This work has been carried out as an activity of the NLO Working Group, 
a collaboration between the Japanese ATLAS group and the numerical analysis 
group (Minami-Tateya group) at KEK.
The author wishes to acknowledge useful discussions with the members, 
especially Y. Kurihara, J. Fujimoto, and A. Bredenstein.


\end{document}